\documentclass[preprint,showpacs,preprintnumbers,amsmath,amssymb]{revtex4}

\usepackage{graphicx}
\usepackage{dcolumn}
\usepackage{bm}
\usepackage{color}
\usepackage{booktabs}
\usepackage{multirow}

\usepackage{threeparttable}

\usepackage{epstopdf}

\usepackage[version=4]{mhchem}

\AtBeginDocument{
	\heavyrulewidth=.08em
	\lightrulewidth=.05em
	\cmidrulewidth=.03em
	\belowrulesep=.65ex
	\belowbottomsep=0pt
	\aboverulesep=.4ex
	\abovetopsep=0pt
	\cmidrulesep=\doublerulesep
	\cmidrulekern=.5em
	\defaultaddspace=.5em
}

\usepackage{hyperref}

\begin{document}

\title{Sensitive detection of metastable NO and \ce{N2} by reactive collisions with laser-excited Li}
\author{Jiwen Guan}
\affiliation{Institute of Physics, University of Freiburg, Hermann-Herder-Str. 3, 79104 Freiburg, Germany}
\author{Tobias Sixt}
\affiliation{Institute of Physics, University of Freiburg, Hermann-Herder-Str. 3, 79104 Freiburg, Germany}
\author{Katrin Dulitz\footnote{Corresponding author: katrin.dulitz@physik.uni-freiburg.de}}
\affiliation{Institute of Physics, University of Freiburg, Hermann-Herder-Str. 3, 79104 Freiburg, Germany}
\author{Frank Stienkemeier}
\affiliation{Institute of Physics, University of Freiburg, Hermann-Herder-Str. 3, 79104 Freiburg, Germany}

\date{\today}

\begin{abstract}
In a proof-of-principle experiment, we demonstrate that metastable nitric oxide molecules, NO(a$^4\Pi_i$), generated inside a pulsed, supersonic beam, can be detected by reactive gas-phase collisions with electronically excited Li atoms in the $2^2$P$_{3/2}$ state. Since the internal energy of NO(a$^4\Pi_i$, $v \leq 4$) is lower than the ionization potential of Li in the $2^2$S$_{1/2}$ electronic ground state, we observe that the product ion yield arising from autoionizing NO(a$^4\Pi_i$)+Li($2^2$S$_{1/2}$) collisions is a factor of 21 lower than the ion yield from NO(a$^4\Pi_i$)+Li($2^2$P$_{3/2}$) collisions. We also compare our findings with measurements of relative rates for collisions of metastable \ce{N2}+Li($2^2$S$_{1/2}$) and metastable \ce{N2}+Li($2^2$P$_{3/2}$) reactive collisions.

Using this detection method, we infer densities of $\approx 600$ NO(a$^4\Pi_i$) molecules/cm$^3$ and $\approx 6 \cdot 10^{4}$ metastable \ce{N2} molecules/cm$^3$ in the interaction region. Our results also allow for an estimate of the fractional population of NO(a$^4\Pi_i$, $v \geq 5$) prior to the collision process. The production of NO(a$^4\Pi_i$) in selected vibrational states using laser excitation from the X$^2\Pi_r$ ground state will open possibilities for the detailed study of vibrational-state-selected NO(a$^4\Pi_i$)-Li($2^2$P$_{3/2}$) collisions.
\end{abstract}

\maketitle

\section{\label{sec:intro}Introduction}
Atoms and molecules in electronically excited states are referred to as metastable, if their decay to lower-lying states is forbidden by electric-dipole selection rules. Owing to their high electronic energy, metastable species can induce the ionization of other atoms or molecules upon collision, a process often referred to as autoionization or, more specifically, as Penning ionization \cite{Siska1993}. Autoionization processes are of great importance in the Earth's upper atmosphere, where metastable species are generated by UV photolysis and electric discharges. For instance, it has been shown that such processes also influence the transmission of radio and satellite signals after thunderstorms \cite{Falcinelli2015}.

Metastable noble gas atoms, such as He(2$^3$S$_1$), He(2$^1$S$_0$) or Ne(3$^3$P$_{0,2}$), are frequently produced and detected in the laboratory \cite{Hotop1996}. Owing to the high efficiency in detecting the ionic reaction products, autoionization studies with metastable noble-gas species often allow for insight into the fundamental nature of chemical reactions \cite{Yencha1978, Siska1993}. For example, recent studies of reactive collisions with metastable noble gases in merged supersonic beams have enabled the observation of quantum resonances and stereodynamic effects \cite{Henson2012, Gordon2017, Gordon2018}.

Metastable states have also been predicted and observed in numerous molecules \cite{Muschlitz1968, Hotop1996}. A summary of selected metastable states of diatomic molecules and their properties is provided in Tab. \ref{tab:metastablemol}. Because of the vast number of possible interaction mechanisms, metastable molecules are particularly interesting for the study of numerous electron and energy transfer processes. Besides autoionization, the stored energy of the metastable molecule can also be released by direct, spin-changing collisional transfer and by a so-called gateway mechanism \cite{Ottinger1994, Ottinger1998}. The latter is a collision-induced energy transfer process to radiating states within a molecule which is mediated by the spin-orbit interaction inherent to the molecule. As this is typically a weak process, it will only operate on pairs of closely spaced levels. The gateway theory has been used to describe the collisional quenching of a number of metastable diatomics, e.g., see Ref.\cite{Ottinger1997} and references therein. Among them, \ce{NO} and \ce{N2} have proven to provide textbook examples of well-defined gateway couplings \cite{Ottinger1994, Heldt1996, Ottinger1997, Ottinger1998}.
\begin{table}[hbt]
	\caption{Natural lifetimes $\tau$ and excitation energies $E^*$ of selected metastable states of diatomic molecules. All state energies, corresponding to the lowest rovibrational state, were taken from Ref. \cite{Huber1979} except for \ce{NO}(a$^4\Pi_i$) \cite{Ottinger1994a, Huber1988}. T = theoretical value, E = experimental value.}
	\label{tab:metastablemol}
	\centering
	\begin{tabular}{lrl}
		\hline
		Species & $E^*$ (in eV) & $\tau$ (in s)\\
		\hline
		\ce{H2}(c$^{3}\Pi_{\textnormal{u}}^{-}$) & 11.762 & 1.02(5)$\cdot$10$^{-3}$ (E) \cite{Johnson1972}\\
		\ce{He_2}(a$^3\Sigma_{u}^+$) & 17.972 & $\approx$ 18 (T) \cite{Chabalowski1989} \\
		\ce{N2}(A$^3\Sigma_{u}^{+}$) & 6.169 & 2.37 (E) \cite{Piper1993}\\
		\ce{N2}(a'$^1\Sigma_{u}^-$)  & 8.398 & 1.7$\cdot10^{-2}$ (T) \cite{Eastes1996}\\
		\ce{N2}(a$^1\Pi_{g}$)        & 8.549 & 120(20)$\cdot10^{-6}$ (E) \cite{Mason1990}\\
		\ce{N2}(E$^3\Sigma_{g}^{+}$) & 11.874& 190(30)$\cdot10^{-6}$ (E) \cite{Borst1971a}\\
		\ce{CO}(a$^3\Pi_1$) & 6.010 & 2.63(3)$\cdot10^{-3}$ (E) \cite{Gilijamse2007}\\
		\ce{NO}(a$^4\Pi_i$) & 4.807 & 0.1 (T) \cite{LefebvreBrion1968}\\	
		\ce{O2}(a$^1\Delta_g$) & 0.976 & 4300(500) \cite{Miller2001}\\
		\ce{O2}(b$^1\Sigma_g^+$) & 1.627 & 12 (E) \cite{Long1973}; 11.65 (T) \cite{Klotz1984}\\
		\hline
	\end{tabular}
\end{table}
Our knowledge about the competitiveness between collision-induced intramolecular energy transfer and interspecies chemical quenching of metastable molecules via autoionization is very limited. The observation of Penning ionization in the \ce{N2}(a$^{'1}\Sigma_{u}^{-}$)+\ce{NH3} system instead of the typical gateway coupling between the radiative C$^{3}\Pi_{u}$ and the metastable a$^{'1}\Sigma_{u}^{-}$ state in \ce{N2} \cite{Ottinger1998} is the only reported case to date.

Owing to its low ionization potential and the accessibility of intermediate states, the NO molecule is one of the most thoroughly studied diatomic molecules. However, very little is known about the nature of the metastable a$^4\Pi_i$ state in NO. Previous results on this state have been indirect and limited because of the lack of efficient NO(a$^4\Pi_i$) production and detection techniques. For instance, this state cannot be efficiently populated by optical excitation from the X$^2\Pi_r$ electronic ground state, as the a$^4\Pi_i \leftarrow$ X$^2\Pi_r$ transition is spin-forbidden. In addition to that, the detection of this state by surface ionization -- a common technique for the detection of metastable noble gases \cite{Hotop1996} -- is very challenging, since the internal energy of this state (4.807 eV \cite{Ottinger1994a, Huber1988} for the lowest rovibrational state) is lower than the work function of most metals.

\begin{figure}[hbt!]  
	\centering
	\includegraphics{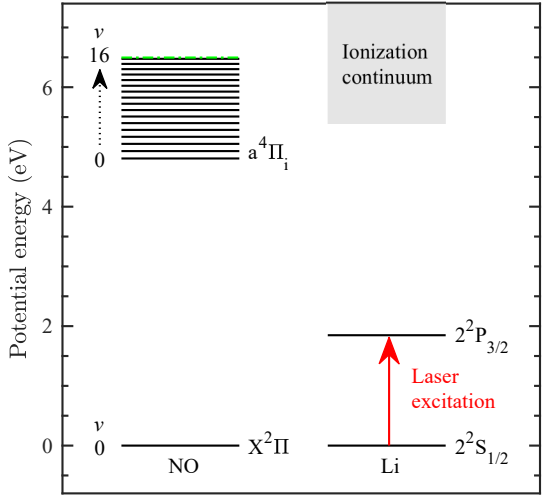}
	\caption{Energy level diagrams of NO and Li highlighting the states important for the present work. Literature values for the term energy and the vibrational constants of the NO(a$^4\Pi_i$) state were taken from Refs. \cite{Ottinger1994a} and \cite{Huber1988}, respectively. The N(2$^4$S$_{3/2}$)+O(2$^3$P$_{2}$) dissociation limit is indicated by a green dash-dotted line \cite{Huber1988}. The Li energies were taken from the NIST Spectral Database \cite{NIST_ASD}.}
	\label{fig:EnergyLevelDiagram}
\end{figure}

In this article, we report on the detection of NO(a$^4\Pi_i$) (``NO$^*$'') by Penning-ionizing collisions with Li atoms. As can be seen from Fig. \ref{fig:EnergyLevelDiagram}, only the higher-lying vibrational states of NO$^*$ have sufficient internal energy to ionize Li atoms from the $2^2$S$_{1/2}$ (``S'') ground state. In contrast to that, the ionization potential of laser-excited Li atoms in the $2^2$P$_{3/2}$ state (``P'') is far below the energy of the lowest rovibrational state of NO$^*$. The comparison of ionization rates for NO$^*$ collisions with ground- and excited-state Li atoms thus allows us to infer additional information about the degree of vibrational excitation in the a$^4\Pi_i$ state of NO. We compare our observations for reactive NO$^*$-Li collisions with results obtained for the reactive scattering of metastable \ce{N2} (``\ce{N2}$^*$'') with Li.

\section{Methods}
Most of the experimental arrangement used in this work has been described elsewhere \cite{Grzesiak2019}. Therefore, only relevant further details are given here. A pulsed supersonic beam of NO is generated by a gas expansion from a room-temperature, high-pressure reservoir (5 bar, purity $>$ 99.5 \%) into the vacuum using a CRUCS valve (24 $\mu$s pulse duration, 7 Hz repetition rate) \cite{Grzesiak2018}. The main valve components are made from stainless steel to prevent a corrosion of the valve by NO gas. Excited NO molecules are produced using a plate discharge source (390 V) attached to the front plate of the valve. A filament (1.45 A, biased to -150 V) is used for injection seeding the gas discharge with free electrons. To eliminate ions and molecules in Rydberg states produced during the discharge, a bias voltage of -300 V is applied to the skimmer which separates the source chamber from the detection chambers. For experiments with \ce{N2}, a beam of \ce{N2} (purity $>$ 99.999 \%) with a backing pressure of 25 bar, a valve pulse duration of 20 $\mu$s and similar discharge conditions as for NO is used. 

The purity of the \ce{NO} and \ce{N2} beams was cross-checked using a quadrupole mass spectrometer. No impurities or other species other than the parent molecules were observed.

As a scattering target, we use an ultracold cloud of $^7$Li atoms confined in a magneto-optical trap (MOT) \cite{Grzesiak2019}. For this, the Li atoms are continuously laser-cooled via the $2^2$P$_{3/2}\leftarrow2^2$S$_{1/2}$ cycling transition at 671 nm in a standard 3D-MOT which is fed from a Li oven via a Zeeman slower. The fluorescence emitted by the Li atoms during the optical cycling process is collected on a CCD camera and used for the calculation of the number of trapped Li atoms. The total number of trapped Li atoms is varied by changing the width of a slit positioned in between the MOT chamber and the Li oven chamber.
In order to achieve a high atomic fraction in the P state, the detuning of the MOT lasers is set to -12 MHz. The laser light for Zeeman slowing is detuned by -60 MHz from resonance and thus does not influence the P state population.

To distinguish between the relative contributions of the S and the P state to the collision process, we have inserted fast mechanical shutters into the MOT and Zeeman laser beam paths. If the shutters are open (closed) when the metastable molecules collide with the Li atoms, a mixture of Li atoms in the S and P states (only Li atoms in the S state) are present. The time window for the observation of the ion production rate is kept short ($< 100 \mu$s, cf. Figs. \ref{fig:TOFs} (a) and (b)). In this way, the spatial expansion of the Li cloud prior to and during the collision process can be neglected in the analysis.

The ions produced by autoionization are collected using an ion time-of-flight (TOF) mass spectrometer with a channel electron multiplier (CEM; Dr. Sjuts Optotechnik GmbH, Standard CEM) built around the center of the Li-MOT. The spectrometer is operated either in continuous or in pulsed mode. The pulsed mode allows for a discrimination of different ion masses. In this case, the voltages applied to the electrodes are simultaneously toggled between high voltage and ground using independent high-voltage switches (Behlke, HTS-41-06-GSM). The continuous mode is used to count the number of ions on the detector. 

\section{Results and Discussion}
\begin{figure}[hbt!]  
	\centering
	\includegraphics{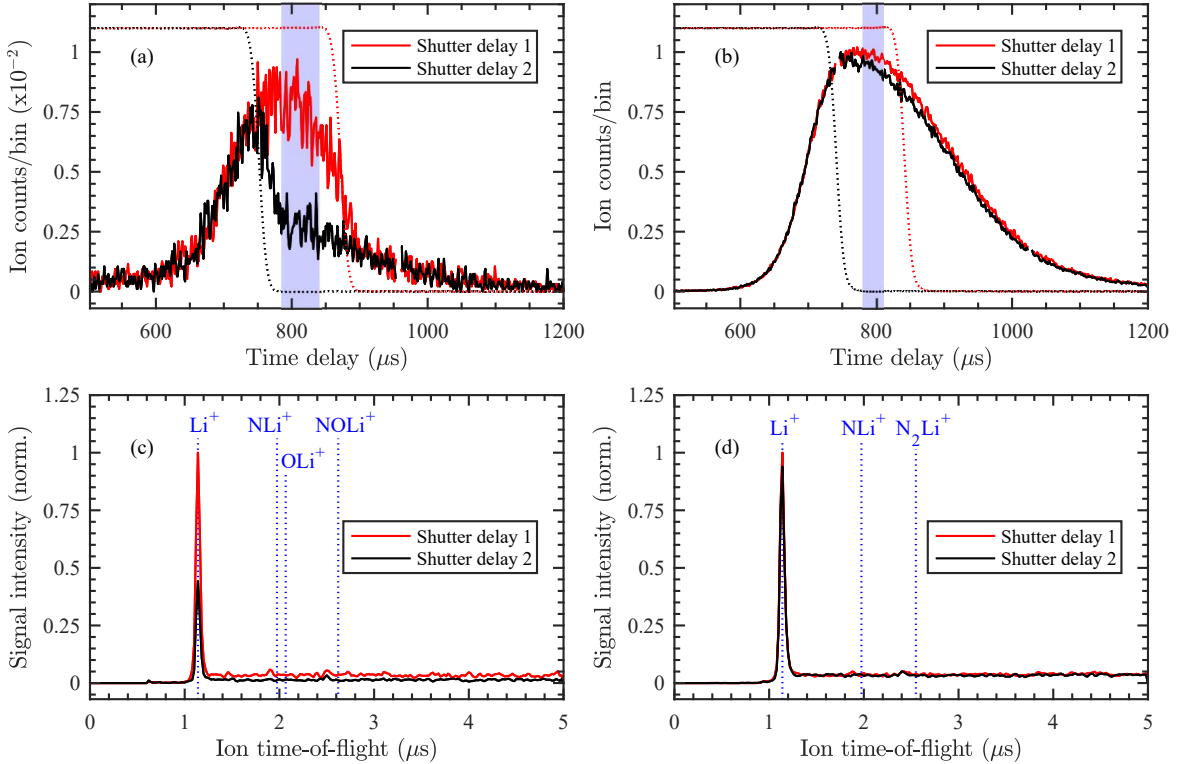}
	\caption{(a) and (b): Measured ion counts (per $1\,\mu$s time bin) as a function of time delay with respect to the valve trigger for (a) NO$^*$-Li collisions and for (b) \ce{N2}$^*$-Li collisions using the continuous detection mode at two shutter time delays. Traces of MOT laser stray light (dotted lines), recorded using a fast photodiode, illustrate the respective shutter closing characteristics. The time windows used for the calculation of ion production rates are indicated by blue-colored shadings. (c) and (d): Normalized ion-time-of-flight traces obtained for (c) NO$^*$-Li collisions and for (d) \ce{N2}$^*$-Li collisions at the same shutter time delays. The expected arrival times of different product ion species are shown as vertical dotted lines. The color code is the same in all four figures.}
	\label{fig:TOFs}
\end{figure}
Fig. \ref{fig:TOFs} shows ion traces for NO-Li and \ce{N2}-Li collisions which were obtained using continuous and pulsed extraction. From the measured ion time profiles in Figs. \ref{fig:TOFs} (a) and (b), we can see that autoionizing collisions are observed $\approx 800\,\mu$s after the valve trigger pulse (and the subsequent electron-impact excitation of \ce{NO} and \ce{N2}), respectively. The widths of the ion profiles are due to the longitudinal spreading of the pulsed supersonic beams from the valve to the interaction region. At times much larger than the laser shutter delay, the shape of the ion profiles is additionally modulated by the decrease in Li atom number which is caused by the spatial expansion of the Li cloud in the absence of a trapping potential. Since a$^4\Pi_i$ is the sole long-lived electronic state in NO, only NO(a$^4\Pi_i$)+Li collisions are expected to occur. In principle, several metastable states of \ce{N2} (cf. Tab. \ref{tab:metastablemol}) and N can be produced by the electron impact excitation of \ce{N2} \cite{Foner1962, Johnson2005, Lofthus1977}. The \ce{N2}(A$^3\Sigma_{u}^{+}$) state has been found to be predominantly populated in a \ce{N2} gas discharge \cite{Foner1962}. Therefore, we expect that the observed \ce{N2}$^*$-Li autoionization signals in our experiment are mainly due to processes involving the A$^3\Sigma_{u}^{+}$ species. 

During the time interval in which the MOT lasers are blocked (cf. dotted lines in Figs. \ref{fig:TOFs} (a) and (b)), only Li atoms in the S ground state are present. From the relative signal intensities in Figs. \ref{fig:TOFs} (a) and (b), it is evident that the NO$^*$-Li(S) autoionization rate is much lower than the NO$^*$-Li(P) autoionization rate. In contrast to that, the observed autoionization rate for \ce{N2}$^*$+Li(S) collisions is only marginally lower than the autoionization rate for \ce{N2}$^*$+Li(P) collisions. Since the experimental conditions were similar in both experiments and the rate coefficients for Penning ionization are expected to be of similar magnitude, we ascribe the observed differences to the energies of the molecular states. The internal energy of all metastable states in \ce{N2}(c.f. Tab. \ref{tab:metastablemol}) is well above the ionization threshold for Li(S) ionization (IP = 5.3917 eV \cite{NIST_ASD}) and for Li(P) ionization (IP = 3.5439 eV \cite{NIST_ASD}). Therefore, both autoionizing \ce{N2}$^*$+Li(S) and \ce{N2}$^*$+Li(P) collisions are energetically allowed. In contrast to that, for NO$^*$-Li(S) collisions, only the energies of NO$^*$ vibrational states with $v \geq 5$ are above the Li(S) ionization threshold (cf. Fig. \ref{fig:EnergyLevelDiagram}). Laser excitation of Li to the P state leads to an effective reduction of the ionization threshold. Therefore, for NO$^*$-Li(P) collisions, all vibrational states of NO$^*$ can contribute to the autoionization of Li.

Autoionizing collisions of a metastable diatomic molecule AB with an atom C can lead to the formation of both atomic and molecular product ions \cite{Yencha1978}:
\begin{equation}
\mathrm{AB^*+C \rightarrow [ABC]^*\rightarrow}
\left\{  
\begin{aligned}
&\mathrm{AB + C^+ + e^-} && \mathrm{Penning\,\,ionization}\\
&\mathrm{ABC^+ + e^-} && \mathrm{Associative\,\,ionization}\\
&\mathrm{AC^+ + B + e^-} && \mathrm{Rearrangement\,\,ionization}\\
&\mathrm{BC^+ + A + e^-} && \mathrm{Rearrangement\,\,ionization}\\
&\mathrm{A + B + C^+ + e^-} && \mathrm{Dissociative\,\,ionization}
\end{aligned}
\right.
\end{equation}
Associative ionization and rearrangement ionization typically play only a minor role in in collisions of metastable rare gases with molecules. For example, in He($2^3$S$_1$)-\ce{H2} collisions, associative ionization and rearrangement ionization amount to only 2 \% and 8 \% of the total ion yield, respectively \cite{West1975}. Since the \ce{ABC+} cations are often weakly bound, they can quickly dissociate if they are formed in vibrationally excited states \cite{Hotop1968}.

Figs. \ref{fig:TOFs} (c) and (d) clearly show that only the products of the Penning ionization process are observed in our measurements. It was postulated that the formation of strongly attractive intermediate states [ABC]$^*$ with ion-pair-state character can make the formation of a bound state of the ABC$^+$ cation less probable \cite{Appolloni1988}. Since Li has a high positive electron affinity \cite{Haeffler1996}, the formation of associative and dissociative ionization products is thus not very likely.

\subsection{Estimate of the metastable molecule densities}

The ions produced by autoionization can be recorded with a high efficiency, which is mainly limited by the characteristics of the detector. In the experiments, we have applied a potential difference of 3.5 kV in between the repeller electrode and the CEM. Under these conditions, the quantum efficiency of the detector for positive ions is \mbox{$\approx$ 50 \%} \cite{DrSjuts}. Therefore, our detection method provides a very sensitive probe for the molecular species contained in a supersonic beam. Since the diameter of the supersonic beam is much larger than the diameter of the Li cloud, the interaction volume is given by the spatial extent of the Li cloud measured using fluorescence imaging. Assuming an interaction volume of 4.6$\cdot 10^{-4}$ cm$^3$ and a typical reaction rate coefficient for Penning ionization at thermal energies, $k \approx 1 \cdot 10^{-9}$ cm$^3$/(molecule s) \cite{Wang1987}, we obtain densities of $\approx 600$ NO$^*$ molecules/cm$^3$ and $\approx 6 \cdot 10^{4}$ \ce{N2}$^*$ molecules/cm$^3$ inside the interaction region. The \ce{N2}$^*$ density is thus a factor of 100 higher than the NO$^*$ density. We believe that this is due to a higher efficiency in producing \ce{N2}$^*$ in an electrical discharge. In addition to that, more than one metastable state of \ce{N2} may be produced in a nitrogen discharge. 

In previous work, NO$^*$ has been detected indirectly by collisional energy transfer to fluorescent states \cite{Ottinger1994, Ottinger1994a, Drabbels1995, Heldt1996, Ottinger1997, Mo1998} and directly by electron emission from a \ce{LaB6} surface and subsequent amplification of the electron signal by a two-channel microchannel plate assembly \cite{Drabbels1995}. Unfortunately, no estimates of the NO$^*$ densities are given by the authors, so that we cannot compare our results with those from previous work. In our laboratory, we have also tried to detect NO$^*$ by secondary electron emission from a freshly Cs-coated metal surface followed by charge amplification. Even though the ionization potential of Cs (3.894 eV \cite{Deiglmayr2016}) is lower then the internal energy of NO$^*$, electron emission by Auger de-excitation was not observed.


\subsection{Relative vibrational populations in NO(a$^4\Pi_i$)}
The residual ion signal in Figs. \ref{fig:TOFs} (a) and (c) for NO(a$^4\Pi_i$)-Li(S) collisions can only be explained by a vibrational excitation of NO(a$^4\Pi_i$) to states with $v \geq 5$. Considerable vibrational excitation of NO$^*$, even in $v \geq 8$,  has also been observed in previous experiments, in which a gas discharge was used for NO excitation \cite{Ottinger1994}.

\begin{figure}[hbt!]  
	\centering 
	\includegraphics{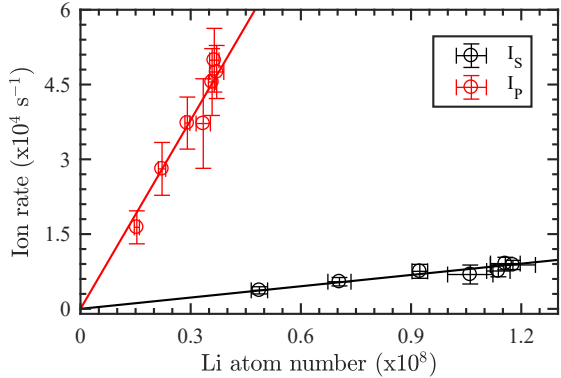}
	\caption{Ion production rates $\dot{I}_{\text{S}}$ and $\dot{I}_{\text{P}}$ for NO$^*$-Li(S) and NO$^*$-Li(P) collisions. The datapoints were derived from the experimental data sets using Eqs. \ref{eq:ionrateMotOn} and \ref{eq:ionrateMotOff}, as a function of Li atom number in the $2^2$S$_{1/2}$ (black markers) and $2^2$P$_{3/2}$ states (red markers), respectively. The red and black lines show curves obtained from weighted linear fits of the respective datasets through the origin. The shown statistical uncertainties ($2 \sigma$) for $\dot{I}_{\text{S}}$ and $\dot{I}_{\text{P}}$ are mainly due to variations of the NO$^*$ density (resulting from shot-to-shot fluctuations of the pulsed valve and the electrical discharge) and the Li density (resulting from changes in laser frequency, polarization and power).}
	\label{fig:Iondensity}
\end{figure}

Since NO$^*$-Li collisions take place within a fixed interaction volume represented by the volume of the ultracold Li cloud, the reaction kinetics can be described by
\begin{align}
\dot{I} = k \cdot N_{\text{NO*}} \cdot N_{\text{Li}}.
\label{eq:GeneralIonProduction}
\end{align} 
Here, $\dot{I}$ is the ion production rate, $k$ is the reaction rate coefficient and $N_{\text{NO*}}$ ($N_{\text{Li}}$) is the number of NO* (Li) particles within the reaction volume, respectively. To apply this equation to our experimental data, two different circumstances have to be considered:
\begin{enumerate}
	\item the time period in which the MOT lasers are turned on, so that a fraction of the Li atoms is optically pumped from the S to the P state
	\begin{align}
	\quad \dot{I}_{\text{on}} &= k_{\text{S}} \cdot N_{\text{NO*}(v \geq 5)} \cdot N_{\text{Li}} \cdot P_{\text{S}} + k_{\text{P}} \cdot N_{\text{NO*}(v \geq 0)} \cdot N_{\text{Li}} \cdot P_{\text{P}}\\
	& \equiv \dot{I}_{\text{S}} \cdot P_{\text{S}} + \dot{I}_{\text{P}} \cdot P_{\text{P}} \label{eq:ionrateMotOn}
	\end{align}
	\item the time period in which the MOT lasers are turned off, so that all Li atoms are in the S state
	\begin{equation}
	\quad \dot{I}_{\text{off}} = k_{\text{S}} \cdot N_{\text{NO*}(v \geq 5)} \cdot N_{\text{Li}} \equiv \dot{I}_{\text{S}} \label{eq:ionrateMotOff}
	\end{equation}
\end{enumerate}
Here $k_{\text{S}}$ ($k_{\text{P}}$) is the rate coefficient for NO*-Li(S) (NO*-Li(P)) autoionization. The fractional steady-state populations of Li in the S and P states are denoted by $P_\text{S}$ and $P_\text{P}$, respectively. From calculations of the optical pumping process using rate equations, we estimate that $P_{\text{S}} = 0.68$ and $P_{\text{P}} = 0.32$ at a MOT laser detuning of -12 MHz, respectively.
Using Eqs. \ref{eq:ionrateMotOn} and \ref{eq:ionrateMotOff}, it is possible to determine the bare ion production rates $\dot{I}_{\text{S}}$ and $\dot{I}_{\text{P}}$ for NO*-Li(S) and NO*-Li(P) collisions, respectively. The results of these calculations are shown in Fig. \ref{fig:Iondensity} for measurements at different Li atom numbers.

Unfortunately, no absolute values for the rate coefficients can be obtained, since the absolute and vibrational-state-dependent NO$^*$ number densities are not known. However, from the slopes of linear fits to the datasets (see Fig. \ref{fig:Iondensity}), we can deduce the ratio $\frac{k_{\text{S}} \cdot N_{\text{NO*}(v \geq 5)}}{k_{\text{P}} \cdot N_{\text{NO*}(v \geq 0)}}$. The energies of all vibrational states of \ce{N2}$^*$ are higher than the ionization potentials of the $2^2$S$_{1/2}$ and $2^2$P$_{3/2}$ states of Li, respectively. Thus, the \ce{N2}$^*$-Li autoionization rates are not vibrational-state-dependent. From Fig. \ref{fig:TOFs} (b), we can infer that the autoionization rates for \ce{N2}$^*$-Li(S) and \ce{N2}$^*$-Li(P) collisions are nearly the same, i.e., $k_{\text{S}} \approx k_{\text{P}}$. Since $N_2$ is a similar diatomic system as NO, we may also assume that $k_{\text{S}} \approx k_{\text{P}}$ for NO$^*$-Li collisions. Thus, the observed change in the ion production rate for NO*-Li collisions can be directly related to the vibrational energy distribution within the molecular beam pulse. In this case, we can estimate that the population ratio $\frac{N_{\text{NO*}(v \geq 5)}}{N_{\text{NO*}(v \geq 0)}} \approx 0.06$ in the interaction zone, i.e. 6 $\%$ of the total NO$^*$ population is located in vibrational states with $v \geq 5$. This fraction corresponds to a vibrational temperature of $\approx$ 2400 K if a Boltzmann distribution is assumed. A comparable vibrational temperature has been measured for NO in the X$^2\Pi_r$ ground state after discharge excitation \cite{Ottinger1994a}.

\section{Conclusion}
In summary, we have investigated a very sensitive scheme for the detection of two metastable molecules, NO(a$^4\Pi_i$) and \ce{N2}$^*$, which is based on collisional autoionization with electronic-state-selected Li atoms. The metastable molecules are produced inside a supersonic gas expansion using a glow discharge. A mixture of Li atoms in the $2^2$S$_{1/2}$ and $2^2$P$_{3/2}$ states is produced by laser cooling in a magneto-optical trap, and electronic quantum-state selection is achieved by rapidly turning off the cooling laser light using mechanical shutters. In all reactive collision systems studied, only the Penning ionization pathway is observed. We do not observe a significant change in the reaction rate for \ce{N2}$^*$+Li($2^2$S$_{1/2}$) vs. \ce{N2}$^*$+Li($2^2$P$_{3/2}$) collisions. In contrast to that, the Li$^+$ product ion yield from NO(a$^4\Pi_i$)-Li collisions markedly increases when the Li atoms are excited to the $2^2$P$_{3/2}$ state. This observation is attributed to the energy-level structure of NO(a$^4\Pi_i$), in which all NO(a$^4\Pi_i$,$v$)-Li($2^2$S$_{1/2}$) collisions are energetically forbidden for $v \leq 4$. We also estimate that approximately $6 \%$ of the NO molecules in the a$^4\Pi_i$ state are in vibrational levels with quantum number $v \geq 5$. No previous information about the relative distributions of vibrational states in NO(a$^4\Pi_i$) is available in the literature.

By using laser excitation instead of an electrical discharge to populate the a$^4\Pi_i$ state from the X$^2\Pi_r$ ground state, it will be possible to selectively excite specific vibrational states in NO \cite{Drabbels1995, Copeland1995, Copeland1997}. Such a state preparation scheme will allow for the detailed study of vibrational-state-selected NO(a$^4\Pi_i$)-Li($2^2$P$_{3/2}$) collisional autoionization and -- in combination with a sensitive fluorescence detector -- its competition with intramolecular collisional energy transfer. It was found that a small external magnetic field ($\leq$ 200 G) shifts the energy levels involved in a gateway process so that energy transfer is inhibited \cite{Ottinger1994, Mo1998}. Therefore, the application of a variable magnetic bias field may even allow for a selective tuning of the relative rates for Penning ionization and energy transfer.

\section*{Acknowledgements}

This work is supported financially by the German Research Foundation (Project No. DU1804/1-1), by the Chemical Industry Fund (Liebig Fellowship to K. Dulitz) and by the University of Freiburg (Research Innovation Fund).

\providecommand{\latin}[1]{#1}
\makeatletter
\providecommand{\doi}
{\begingroup\let\do\@makeother\dospecials
	\catcode`\{=1 \catcode`\}=2\doi@aux}
\providecommand{\doi@aux}[1]{\endgroup\texttt{#1}}
\makeatother
\providecommand*\mcitethebibliography{\thebibliography}
\csname @ifundefined\endcsname{endmcitethebibliography}
{\let\endmcitethebibliography\endthebibliography}{}

\end{document}